\documentclass[preprint,showpacs,preprintnumbers,amsmath,amssymb]{revtex4}

\begin{document}

\preprint{Phys. Rev. E {\bf 69}, in press (2004).}

\title{Eigen Model as a Quantum Spin Chain: Exact Dynamics}

\author{David Saakian$^{1,2}$}
\author{Chin-Kun Hu$^{1}$}
\email{huck@phys.sinica.edu.tw}
\affiliation{$^1$Institute of Physics, Academia Sinica, Nankang, Taipei 11529, Taiwan}
\affiliation{$^2$Yerevan Physics Institute,  Alikhanian Brothers St. 2,
Yerevan 375036, Armenia }

\date{\today}

\begin{abstract}
We map Eigen model of biological evolution [Naturwissenschaften  {\bf 58}, 
    465 (1971)] into a one-dimensional quantum spin model 
with non-Hermitean Hamiltonian.  Based on such a connection, we derive exact
relaxation periods for the Eigen model to approach static energy landscape
from various initial conditions. We also study a simple case of dynamic fitness
function.
\end{abstract}
\maketitle

\vskip 5 mm


Eigen model of asexual evolution 
\cite{eigen71,eigen89} is one of the main mathematical models
in this field. In this model  individuals  have offsprings, that are 
subjected to mutation that connects with a selection rule. 
In his original work Eigen found an error threshold  similar to the
critical point in critical phenomena such that when the mutation is larger
than the error threshold  the organism can not survive. 
Later, statistical mechanics
has been applied to investigate the discrete time version 
of the original model \cite{leut87,tara92}.
Franz and Peliti \cite{fp97} 
derived another important result in the Eigen model: concentration
of individuals  around the peak configuration.

In the parallel mutation-selection model, an alternative to the 
Eigen model, a mutation mechanism and 
 a selection mechanisms are two independent processes that take 
 place concurrently \cite{ck70}.
Baake {\it et al.} \cite{bbw97} proved that for the 
parallel mutation-selection scheme, the time evolution 
equation for the frequencies of different species is equivalent 
to the Schr\"odinger equation in imaginary time for quantum 
spins in a transverse magnetic field. 
Based on such a connection, recently we used 
Suzuki-Trotter  formalism  \cite{Su} to study
both statics and dynamics of  the  model with a single peak fitness function
 \cite{sh}.
In the present Letter, we will extend such study to the
Eigen model \cite{eigen71} by reexpressing  the Eigen model's dynamics 
via  quantum chain problem, then solving  the dynamics to obtain
exact relaxation periods for the Eigen model.  
The dynamic aspects  play important role 
during the evolution in changing environments 
 \cite{gill91,Ohta,wrm01}. Thus such aspects in the Eigen model
 have been considered in recent  works \cite{kb02,sn00}, in which  approximate 
 formulas for the  relaxation periods have been found and
applied  to describe a virus-immune system coevolution. 
Our equations for exact relaxation periods are consistent
with approximate formulas in  Refs. \cite{kb02,sn00} for the case
of  one mutation per replication.

As in Ref. \cite{sh}, the  genome 
configuration is specified by a sequence of $N$ spin values  
 $s_k=\pm 1$, $1 \le k \le N$. 
We denote the $i$-th genome configuration by 
$S_{i}\equiv{(s_{1},s_{2},...,s_{N})}$ and 
the probability of the $i$-th
genome at time $t$ is given by $p_{S_i}\equiv p_i(t)$ and
the fitness $r_{i}$ is the average number of offspring's per 
unit time. 
In our language, the chosen fitness $r_i$ is a  function $f$ 
that operates on the genome configuration 
$S_{i}$, i.e., $r_{i}=f(S_{i})$. 

In the Eigen model, elements of the 
 mutation matrix $Q_{ij}$  represent the probability that an offspring 
produced by state $j $
 changes to state $i$, and the evolution is 
given by the set of $2^N$ coupled equations for $2^N$ probabilities $p_{i}$ 
\begin{equation}
\label{e1}
\frac{dp_i}{dt}= \sum_{j=1}^{2^N}Q_{ij} r_j  p_j-p_i(
\sum_{j=1}^{2^N}r_{j}p_j).
\end{equation}
Here  $p_i$ satisfies $\sum_{i=1}^{2^N}p_i=1$ and 
$Q_{ij}=q^{N-d(i,j)}(1-q)^{d(i,j)}$ with 
$ d(i,j)\equiv (N-\sum_{l=1}^Ns^l_is^l_j)/2$  being the Hamming 
distance between  $S_i$ and $S_j$. The parameter $1-q$ describes the 
efficiency of mutations. 
 For the parallel mutation-selection model,
 the dynamics is given by
 \begin{equation}
 \label{e2}
 \frac{dp_i}{dt}=\sum_{j=1}^{2^N} m_{ij}p_j
   +p_ir_i-p_i (\sum_{j=1}^{2^N} r_jp_j),
 \end{equation}
 where $m_{ij}$ are the elements of the mutation matrix 
$m_{ij}=\gamma_0$ for $d(i,j)=1$, $m_{ij}=-N\gamma_0$ for $i=j$,
and $m_{ij}=0$ for $d(i,j) > 1$.

Eigen found that it is enough to solve Eq.  (\ref{e1})
 for only linear parts \cite{eigen71}. 
Let us decompose the first, linear  part of Eq. (\ref{e1})  
via mutations to the fixed length  $d(i,j)=l$:
\begin{equation}
\label{e3}
\frac{dp_i}{dt}= \sum_{l=0}^{N}\sum_{j,d(i,j)=l}Q_{ij} r_j  p_j. 
\end{equation}
The second sum is over all configurations having Hamming 
distance $l$ from the peak configuration. 
Using the relation $\sum_{i=1}^{2^N} Q_{i,j}=1$, we can show
that when $p_i$ satisfies Eq. (\ref{e3}), then
 \begin{equation}
 \label{e4}
p_{i}^{\prime}(t) \equiv \frac{p_i(t)}{\sum_jp_j(t)}
 \end{equation}
satisfies Eq. (\ref{e1}).
We can compare Eq. (\ref{e3}) with Eq. (\ref{e2})   without the last nonlinear term. 
The terms $l=1$ and $l=0$ in Eq. (\ref{e3})  correspond, respectively,  to the first and second
terms in Eq. (\ref{e2}). In  Eq. (\ref{e3}),  there are terms with higher level $l\ge 2$ spin flips.
Baake {\it et al.}  \cite{bbw97} mapped   Eq. (\ref{e2})  into a model of quantum spin chain.
Here we will  use the same method to map the model of   Eqs. (\ref{e1}) and (\ref{e3}) 
 into a quantum spin model  with additional higher level spin flip terms.

Let us reformulate the system  of Eq. (\ref{e3}).  As we identify configuration $S_j$ with 
a collection of spins $s^j_1..s^j_N=\pm 1$ and 
define fitness function $f$ as $r_j=f(s^j_1..s^j_N)\equiv f(S_j)$.  Let us consider
 vectors in the Hilbert space of  $N$  quantum
 Pauli spins. With the $p_i$ of  Eq.(\ref{e3}),  we connect a vector in Hilbert
 space  $\sum_{i=1}^{2^N}p_i|S_i>$.  Then $r_j\to f(\sigma^z_1..\sigma^z_N)$.
The  $l$ spin flip term $Q_{ij}$ in Eq. (\ref{e3})  can be identified with 
 a matrix element $<S_j|D_l|S_i>$ of quantum operator
 \begin{equation}
 \label{e5}
D_l\equiv q^{N-d(i,j)}(1-q)^{d(i,j)}\sum_{1\le i_1<..i_l\le N}\sigma^x_{i_1}..\sigma^x_{i_l}.
\end{equation}
 Thus Eq. (3) is equivalent to Scr\"odinger equation:
 \begin{eqnarray}
 \label{e6}
  -H=f(\sigma^z_1..\sigma^z_N)q^N
+ q^N\sum_{l=1}^N(\frac{1-q}{q})^l\sum_{(1\le i_1<i_2..i_l\le N)}^{2^N}\sigma^x_{i_1}..\sigma^x_{i_l}
f(\sigma^z_1..\sigma^z_N),\nonumber\\
 \frac{d}{dt}\sum_{j=1}^{2^N}p_j(t)|S_j>=
 -H\sum_{j=1}^{2^N}p_j(t)|S_j>
\end{eqnarray}
and Eq.(4) to:
 \begin{eqnarray}
 \label{e7}
 &&Z=\sum_{ij}<S_i | e^{-Ht}|S_j>p^0_j\nonumber\\
  &&p_i=\frac{\sum_j  <S_i | e^{-Ht}|S_j>p^0_j }{Z},
 \end{eqnarray}
 where $\sigma$ denotes the spin operator and $|S>$ is the standard 
 notation for the spin state. 
One can multiply Eq. (6) from the left by $<S_i|$ and obtain
Eq. (\ref{e3}).

For the single-peaked fitness function,  we take 
 \begin{eqnarray}
 \label{e8}
f(S_1)=A, {~\rm and~}
f(S_i)=1  ~{\rm for}~ i\ne 1, 
 \end{eqnarray}
with
 $S_{1}\equiv(+1,+1,...,+1)$, which is
equivalent to choosing 
 \begin{eqnarray}
 \label{e9}
f(S_1)=1+(A-1)[\frac{\sum_{i=1}s_i}{N}]^p
 \end{eqnarray}
at the limit $p \to \infty$.
A careful look at the Hamiltonian of Eq. (\ref{e6})  reveals  that it is non-Hermitean.
But we will mainly work with the matrix elements between $S_i \ne S_1$ and $S_j\ne S_1$
and  for these situations we can miss the multiplier 
$f(\sigma_1^z..\sigma_N^z)=1$.
For that sector of Hilbert space,  Hamiltonian is Hermitean.
To investigate the dynamics,  we are using the matrix elements of Hamiltonian
 \begin{eqnarray}
 \label{e10}
-<S_1|H|S_1>=Ae^{-\gamma};& \nonumber\\
<S_i|H|S_j>=<S_i|H_{diff}|S_j>,~i  \ne 1;& \nonumber\\
-H_{diff}=\hat Ie^{-\gamma}+
\sum_{l=1}^Ne^{-\gamma}(\frac{1-q}{q})^l\sum_{1\le i_1
<i_2..i_l\le N}^{2^N}\sigma^x_{i_1}..\sigma^x_{i_l}&,
 \end{eqnarray}
where $\hat I$ is identity operator, $\gamma\equiv -N\ln (q) \approx N(1-q)\ll N$. For us  only terms
 $l\ll N$ are relevant, therefore 
the substitution $q^N[(1-q)/{q}]^l\to e^{-\gamma}(\gamma/N)^l$ can be applied.

To calculate matrix elements of $T(t)\equiv e^{- Ht}$,  one should introduce the 
Suzuki-Trotter formalism \cite{Su}. To perform analytical calculation,  it is more convenient
to use Eq.  (\ref{e9})  for the fitness function and   Eq. (\ref{e10}).
For any value of $p$ an exact method of Suzuki-Trotter formalism  \cite{Su}
can map the system to the  problem in classical statistical mechanics. 
Moreover, for the large values of $p$ it is well known  that 
the problem is drastically simplified.  
For the  quantum $p$-spin interactions in a transverse magnetic field,
Goldschmidt \cite{yyg} has found
that all the order parameters (magnetizations) are either $1$ or $0$ and  
one should take either only transverse interaction terms 
($\sigma^x_{i_1}..\sigma^x_{i_l}$ ) or only the longitudinal one
 ( $e^{-\gamma}[1+(A-1)({\sum_i \sigma_i ^z}/{N})^p])$. 
Therefore,  we can work with system of  Eq. (\ref{e10}) using the following trick. With
 exponential accuracy of order $1/2^N$, 
 it is possible to neglect  the $\sigma^x_i$ terms  
in Eq. (\ref{e6}) and get 
 \begin{equation}
 \label{e11}
 <S_1|e^{-Ht}|S_1> \sim \exp[(Ae^{-\gamma})t].
 \end{equation}
Matrix elements $<S_i|e^{-Ht}|S_j>$ for $i \ne 1$ 
can be replaced with exponential  accuracy 
 by  $<S_i|\exp[-H_{diff}t]|S_j>$. Equation 
 \begin{equation}
 \label{e12}
 \frac{d}{dt}\sum_{i=1}^{2^N}x_i(t)|S_i>=-H_{diff}\sum_{i=2}^{2^N}x_i(t)|S_i>
 \end{equation}
is equivalent to  Eq. (\ref{e3}) with $r_j=1$ for $j=2, \dots, 2^N$ and 
$r_1=0$.  Then we derive that 
 \begin{equation}
 \label{e13}
 \sum_{i=2}^{2^N}x_i(t)=\exp[t]\sum_{i=2}^{2^N}x_i. 
 \end{equation}
From Eqs.  (\ref{e11}) and  Eq. (\ref{e13}), we have   $p_1\sim \exp[(Ae^{-\gamma})t]$  and 
$\sum_{i=2}^{2^N}p_i\sim e^t$.
Therefore,  we derive the Eigen's exact formulae for the error threshold,
 \begin{equation}
 \label{e14}
 A>e^{\gamma}.
 \end{equation}

Let us calculate now the transition probabilities
$<S_j|\exp[-H_{diff}t]|S_i>$  between two states with the total number of $M$ flips
between configurations $S_i\equiv\{s^i_1..s^i_N\}$ and $S_j\equiv\{s^j_1..s^j_N\}$
and define $m=1-2M/N$.
We will show below that the model can be solved at
\begin{equation}
\label{e15}
\frac{1}{N}\sim (1-q)\ll 1.
\end{equation}
For the finite $(1-m)$,  we guess that the relaxation time $t$ is of order $N$ and define
\begin{equation}
\label{e16}
T=te^{-\gamma}/N. 
\end{equation}
There are $N(1+m)/2$ spins without flips (+1 spins) and $N(1-m)/2$  flipped spins (-1 spins). 
Let us denote by $h_l$ the term of $l$ spin flip in the Hamiltonian.  To calculate the matrix 
element $<S_j|\exp[-H_{diff}t]|S_i>\equiv <S_j|\exp[-t\sum_lh_l]]|S_i>$,  let us use an equality
$\exp[a\sigma^x_{i_1}\sigma^x_{i_2}\dots\sigma^x_{i_l}]=\cosh[a][1+\tanh[a]
\sigma^x_{i_1}\sigma^x_{i_2}\dots\sigma^x_{i_l}]$ and expand the product
keeping terms till  the $M$-th degree: 
\begin{eqnarray}
\label{e17}
<S_j|e^{-tH_{diff}}|S_i>\approx 
\sum_{K=1}^M\sum_{l_1+..l_K=M}\frac{M!}{l_1!l_2!..}\nonumber\\
\cosh(\gamma T)^N\tanh(\gamma T)^{l_1}
\prod_{i>1}[\frac{(T\gamma^i)<+|\sigma^x_1|->}{N^{i-1}i!}]^{l_i}. 
\end{eqnarray}
We find via the saddle point the principal term in the expression of Eq. (17)
among all distributions with different $l_i$.
We keep $\cosh,\tanh$ only for the one spin
flip terms.  We calculate also the combinatorics of insertion into
$M$  site box combination of $l_1$ single points, $l_2$ duplets,...$l_k$
$k$ plets, which satisfy the constraint 
\begin{equation}
\label{cons}
\sum_{i=1}^Mil_i=M.
\end{equation}


We can take the constraint of Eq.(\ref{cons}) into account via a 
Lagrange parameter $\lambda$
 and write $l_i$ as $x_iN$.
For the logarithm of a typical term for summation in Eq. (17), we have
\begin{eqnarray}
\label{e19}
N\phi(T,m,\gamma)\equiv N[ \ln \cosh(\gamma T)
+x_1\ln (\tanh(\gamma T))\nonumber\\
+\frac{1-m}{2}\ln \frac{1-m}{2}-
\frac{1-m}{2}- \sum_{i=2}(x_i\ln (x_ii!/T)-x_i)\nonumber\\
+\ln \gamma \sum_{i=2}^Mix_i-x_1\ln x_1+x_1+\lambda(\sum_iix_i-\frac{1-m}{2})]. 
\end{eqnarray}
 The extremum conditions for $x_i$ of Eq. (\ref{e19}) give:
\begin{eqnarray}
\label{e20}
x_1=\tanh(\gamma T)z/\gamma,~ i!x_i=Tz^i,i\ge 2, 
\end{eqnarray}
where $z\equiv \gamma e^{\lambda}$.  Using  formulas:
$ \sum_{i=2}^Mx_i=T\sum_{i=2}{z^i}/{i!}=T(\exp (z)-z-1)$,
$\sum_{i=2}^M {iz^i}/{i!}=\sum_{i=1}^M {iz^i}/{i!}-z= z\exp (z)z-z$,
$\sum_{i=2}^Mx_i\ln (x_ii!/T)=T\ln z\sum_{i=2}{iz^i}/{i!}=
Tz \ln z(\exp(z)-1)$,
 and Eq. (\ref{cons}), we have:
\begin{eqnarray}
\label{e21}
zTe^z-Tz +z\tanh (\gamma T)/\gamma=\frac{1-m}{2}, \nonumber\\
\phi(T,m,\gamma)=\frac{1-m}{2}\ln \frac{(1-m)\gamma}{2}-\frac{1-m}{2}\nonumber\\
+\ln \cosh(\gamma T)+
z\tanh(\gamma T)[1-\ln z]/\gamma \nonumber\\
+T[e^z(1-z\ln z)-z(1-\ln z)-1].
\end{eqnarray}

Let us now consider an ansatz for $<S_1 |e^{-Ht}|S_i>$:
\begin{eqnarray}
\label{e22}
<S_1|\exp [AN(T-T_0)]|S_1> <S_1|e^{-H_{diff}t_0}|S_i> \nonumber\\
=\exp\{N[ A(T-T_0) + \phi(T_0,m,\gamma)]\}.
\end{eqnarray}
While calculating this expression via saddle point, we first find the extremal point 
$T_0\equiv e^{- \gamma}t_0/N$ from the saddle point condition:
\begin{eqnarray}
\label{e23}
A=\frac{d\phi(T_0)}{dT}. 
\end{eqnarray}
The transition period $t_1\equiv Ne^{\gamma}T_1$ is defined from the condition, that the contribution 
$<S_1|e^{-Ht}|S_i>$  into $Z$ of Eq. (7) is larger than the contributions of other terms
$<S_j|e^{-Ht}|S_i>$ (equal to  $e^t$ according to Eq. (13)):
\begin{eqnarray}
\label{e24}
exp({N[\phi(T_0,m,\gamma)+A(T_1-T_0)]}) \ge exp({Ne^{\gamma}T_1}), \nonumber\\
T_1=\frac{A}{A-e^{\gamma}}T_0-\frac{\phi(T_0,m,\gamma)}{A-e^{\gamma}}.
\end{eqnarray}
Thus Eqs. (21), (23)-(24)  give the relaxation period 
$T_1\equiv e^{-\gamma}t_1/N$  under the constraint of Eq. (\ref{e14})  for the  fitness $A$.

There are several phases in dynamics. For $0<t<t_0$,  there is a random drift
to the peak configuration $S_1$. For $t_0<t< t_1$,
there is a growth in the value of  $p_1$, but the macroscopic majority is still out of the peak 
configuration. For $t>t_1$, the  macroscopic majority is near the peak configuration.

Let us give an explicit expressions for the case 
\begin{equation}
\label{e25}
\frac{\gamma (1-m)}{A}\ll 1. 
\end{equation}
 This is a typical biological  situation for observing  $1-m\ll 1$. 
In this case, as we can check later,  $T\sim (1-m)\ll 1$, thus  one can replace 
$z\tanh(\gamma T)/\gamma\to z T $ and derive a simplified
system of equations:
\begin{eqnarray}
\label{e26}
\phi(T,m,\gamma)&=& \frac{1-m}{2}[\ln \gamma\frac{1-m}{2}-(1+\ln z)]
+T(e^z-1), \nonumber\\
Tze^z &=&\frac{1-m}{2}, \nonumber\\
\frac{d\phi}{dT}&= & e^z-1=A. 
\end{eqnarray}
Then $T_0=(1-m)/[ 2(1+A)\ln (1+A)]$.
Thus for the relaxation period $t=T_1e^{\gamma}N$,  one has an expression:
\begin{equation}
\label{e27}
t_1= (1-m)N\frac{\ln \frac{2e\ln (A+1)}{(1-m)\gamma}}{2(Ae^{-\gamma}-1)}.
\end{equation}
Equation  (\ref{e27})  gives relaxation period from the original distribution, 
concentrated at the configuration with the overlap $Nm$ with the peak
fitness configuration, and mutation per site $1-q=\gamma/N$. The physical
 meaning of the term $\frac{(1-m)N}{2}$ is trivial (for the case of 
infinite population): the relaxation period is proportional to the 
Hamming distance. We can understand also the term
$(Ae^{-\gamma}-1)$ in the dominator: it is a natural consequence 
of the fact that relaxation period should diverge
at the error threshold $Ae^{-\gamma}\to 1$.
 Our derivation is valid when the condition of  Eq. (\ref{e25})  is satisfied. 
 An  estimate for the $t_1$ has been given in Refs. \cite{sn00,kb02}.
\begin{equation}
\label{e28}
t_1=\frac{\ln \frac{1}{1-q}}{Ae^{-N(1-q)}-1}
\equiv\frac{\ln \frac{N}{\gamma}}{Ae^{-\gamma}-1}. 
\end{equation}
We note that Eq. (\ref{e28}) is qualitatively correct and 
consistent with Eq. (\ref{e27}) for the case $N(1-m)/2=1$ 
considered in that works.
Our derivation is rigorous only for a large number
of flipped spins, i.e. $N(1-m)/2 >>1$.  For a small number of flipped spins
considered in Refs. \cite{sn00,kb02}, we still can not
derive an exact analytical formula.

Let us briefly consider a simple case  of  a dynamic fitness landscape:
a fitness peak $A(t)$ in the first configuration $S_1$,  which  changes with the time.
Now for the $<S_1|e^{-Ht}|S_1>$,   we have $\exp[e^{-\gamma}\int_0^t A(\tau){\it d} \tau]$.
Equations  (\ref{e23})  and (\ref{e24}) transform  into 
\begin{equation}
\label{e29}
A(\tau_0)=\frac{d\phi(T_0)}{dT_0},
~~\phi(T_0,m,\gamma)+\int_{T_0}^{T_1}A(\tau){\it d} \tau]>e^{\gamma} T_1.
\end{equation}
Now could be a very rich phase structure with different solutions for $T_0$.
For the $T_1\equiv t_1e^{-\gamma}/N$, we have:
\begin{eqnarray}
\label{e30}
\hat A=\frac{\int_{T_0}^{T_1}A(\tau)d \tau}{T_1-T_0},~~
T_1=\frac{\hat A}{\hat A-e^{\gamma}}T_0-\frac{\phi(T_0,m,\gamma)}{\hat A-e^{\gamma}}.
\end{eqnarray} 
Now $A$ is replaced with a mean value. For the case 
of $A\gg \gamma (1-m)$, we again have Eq. (\ref{e27}),  only with $A\to \hat A$.

For $A\gg 1$, we can calculate  the relaxation time from
an original uniform distribution on a static landscape: $p_i=1/2^N$.
For this purpose, we compare the contribution 
$<S_1|e^{-Ht}|S_1>=
{2^{-N}} \exp[Ae^{-\gamma}t]$ with  $\exp(t)$ (sum of other contributions) 
for their contributions to $Z$ of Eq. (\ref{e7}) and find that
\begin{equation} 
t_1= \frac{N\ln 2}{Ae^{-\gamma}-1}.
\end{equation}

To derive the steady state distributions of $p_i$, we can set $dp_i/dt=0$ in 
Eq. (\ref{e1}).
 For $A\gg 1$ we can derive
that $p_i=q^N[(1-q)/{q}]^{d(1,i)}$ and
and the result obtained 
in Ref. \cite{fp97}: 
$\frac{1}{N}\sum_ip_i\sum_{l=1}^N s_i^l=2q-1.$.

Let us briefly consider the case of two isolated  flat peaks in fitness
 landscape with fitness heights $A_1$ and $A_2$, 
and widths $g_1$ and $g_2$. The peak of height $A_i$
has $g_i$ one-flip neighbors of the same height.
A simple consideration gives for the effective fitness $A_i[1+g_i(1-q)]$. 
Thus the Svetina-Scuster phenomenon \cite{ss}
for two peaks appears at $A_1[1+g_1(1-q)]=A_2[1+(1-q)g_2]$.

In 1971,  Eigen \cite{eigen71} found an exact error threshold for
his model from information theory arguments. After more than 30 years of different
approximate or numerical investigations of the Eigen model,
we have found the exact dynamics of the model presented in Eqs. (21), (23), and (24).
Our Eq. (27) gives the relaxation periods with a high degree of accuracy $O (1-m)^2\sim (d/N)2$, it is 
more accurate  than Eq.(28) derived in \cite {sn00,kb02}.
In \cite{sh} we compared the accurate result of this work Eq.(27) with the corresponding relaxation period
of parallel scheme to conclude, that even at the limit of vanishing mutation rates two mutation schemes
give a finite (nonvansihing) difference in relaxation periods. Therefor there is at least one situation,
when our exact  Eq.(24) or accurate approximation Eq.(27) give new qualitative result.
We have also applied the similar method to study a simple case of dynamical environments
and obtained  Eqs. (\ref{e29}) and (\ref{e30}). 
The more involved situations with a very rich and 
interesting phase structure \cite{wrm01}
as well as the virus-immune system coevolution \cite{kb02}
can  also be investigated by our method.
The main open problem is an application of the 
same method to the finite population case. In this case the search of a a peak configuration could
be exponentially large function of $N$, instead of a linear in Eq. (27).
We hope that progress in this direction is possible in the near future, considering funnel like
fitness landscapes. In any case in this work we considered Eigen model's dynamics as a some statistical
mechanics problem and exactly solve it. 

This work was partially supported by the National Science Council of 
 the Republic of China (Taiwan) under Grant No. NSC 92-2112-M-001-063.


\begin{thebibliography}{30}




 \bibitem{eigen71} M. Eigen, Naturwissenschaften  {\bf 58}, 
    465 (1971).

 \bibitem{eigen89} M. Eigen, J.  Mc  Caskill, and P. Schuster,  Adv.  Chem. Phys.
           {\bf 75}, 149 (1989).

 \bibitem{leut87} I. Leuthausser,  J. Stat. Phys.   {\bf 48}, 343  (1987).

 \bibitem{tara92} P. Tarazona, Phys. Rev. A  {\bf 45}, 6038 (1992).

 \bibitem{fp97} L. Franz and L. Peliti,  J. Phys. A. 
  {\bf 30}, 4481 (1997).

 \bibitem{ck70} J. F.  Crow and  M. Kimura, {\it  An Introduction to Population 
  Genetics Theory}  (Harper Row, New York, 1970).

 \bibitem{bbw97} E. Baake, M. Baake, and  H. Wagner, Phys. Rev. Lett. 
   {\bf 78}, 559 (1997).

\bibitem{Su} M. Suzuki, {\it Quantum Monte Carlo Methods}
(Springer Verlag, Berlin, 1986).

\bibitem{sh} D. B. Saakian and C.-K. Hu, Solvable Biological Evolution Model 
 with Parallel Mutation-selection  Scheme, submitted to PRE.


\bibitem{gill91} J. H. Gillespie, {\it The Cause of Molecular Evolution} 
                                 (Oxford Univ. Press, Oxford, 1991).


\bibitem{Ohta} T. Ohta, Proc. Nat. Acad. Sci. USA {\bf 90}, 10676 (1993).

 \bibitem{wrm01} C.  O.  Wilke, C. Ronnewinkel,and  T. Martinetz, Phys. 
    Rep.  {\bf 349}, 395 (2001).


\bibitem{sn00}M.Snoad, N.Snoad, Phys.Rev.Lett. {\bf 84}, 191 (2000).

\bibitem{kb02}C.Kamp, S.Bornholdt, Phys.Rev.Lett. {\bf 88},068104(2002).





\bibitem{yyg} Y. Y.  Goldschmidt, Phys. Rev.  {\bf B41}, 4858(1990).


\bibitem{ss}P.Schuster,J.Swetina, Bull. Math. Bio. {\bf 50},635(1988). 





\end{thebibliography}
 \end{document}